# Experimental evidence of anomalously large superconducting gap on topological surface state of $\beta$-Bi$_2$Pd film

J.-Y. Guan[1,2,*], L.-Y. Kong[1,2,*], L.-Q. Zhou[1,2], Y.-G. Zhong[1,2], H. Li[1,2], H.-J. Liu[1,2], C.-Y. Tang[1,2], D.-Y. Yan[1,2], F.-Z. Yang[1,2], Y.-B. Huang[3], Y.-G. Shi[1,5], T. Qian[1,4,5], H.-M. Weng[1,5], Y.-J. Sun[1,4,5,‡] and H. Ding[1,4,5,†]

[1] Beijing National Laboratory for Condensed Matter Physics and Institute of Physics, Chinese Academy of Sciences, Beijing 100190, China
[2] School of Physics, University of Chinese Academy of Sciences, Beijing 100190, China
[3] Shanghai Synchrotron Radiation Facility, Shanghai Institute of Applied Physics, Chinese Academy of Sciences, Shanghai 201204, China
[4] CAS Center for Excellence in Topological Quantum Computation, University of Chinese Academy of Sciences, Beijing 100190, China
[5] Songshan Lake Materials Laboratory, Dongguan, Guangdong 523808, China

**Connate topological superconductor (TSC) combines topological surface states with nodeless superconductivity in a single material, achieving effective *p*-wave pairing without interface complication. By combining angle-resolved photoemission spectroscopy and *in-situ* molecular beam epitaxy, we studied the momentum-resolved superconductivity in $\beta$-Bi$_2$Pd film. We found that the superconducting gap of topological surface state ($\Delta_{TSS}$ ~ 3.8 meV) is anomalously enhanced from its bulk value ($\Delta_b$ ~ 0.8 meV). The ratio of $2\Delta_{TSS}/k_B T_c$ ~ 14.2, is substantially larger than the BCS value. By measuring $\beta$-Bi$_2$Pd bulk single crystal as a comparison, we clearly observed the upward-shift of chemical potential in the film. In addition, a concomitant increasing of surface weight on the topological surface state was revealed by our first principle calculation, suggesting that the Dirac-fermion-mediated parity mixing may cause this anomalous superconducting enhancement. Our results establish $\beta$-Bi$_2$Pd film as a unique case of connate TSCs with a highly enhanced topological superconducting gap, which may stabilize Majorana zero modes at a higher temperature.**

The studies of superconducting (SC) topological surface states [1] have been propelled by the prospect of harboring vortex-confined Majorana zero mode (MZM) [2,3], which is widely believed to be a building block of fault-tolerant quantum computation [4]. Theoretically, MZMs can emerge as a special type of Bogoliubov excitations in an intrinsic topological superconductor (TSC) with *p*-wave pairing [5,6], or in artificial designs combing conventional *s*-wave superconductivity with special band structures [1,7–27], *e. g.,* the topological insulating states [1, 7–9,16,20–22]. In the latter case, a superconducting topological surface state has been proved to play a similar role as a two-dimensional *p*-wave superconductor [1,2]. An effective *p*-wave superconductivity can be realized on the interface of a proximitized heterostructure between an *s*-wave superconductor and a strong topological insulator (TI) [16,20–22], or on the sample surface of a self-proximitized connate TSC [28–53], *i. e.,* a full-gapped bulk superconductor holds topological surface states [47,51]. Heterostructures usually suffer shortcomings such as gap softness [54,55] and fragile device fabrication [26,27], thus are difficult for observing and manipulating MZMs in experiments [55,56]. It has been a long sought-after goal to find an ideal platform which can easily create, measure and control MZMs.

Recently, topological surface states and MZMs are observed clearly in single material platforms of Fe(Te,Se) bulk single crystals [38–50] and similar compounds of iron-based superconductors [49,51–53], which indicates that a connate TSC is a promising platform for pursuing topological quantum computation [47].

The SC gap of topological surface state ($\Delta_{TSS}$) plays a vital role in protecting MZM that a larger $\Delta_{TSS}$ leads to a larger energetic separation between MZM and other trivial excitations [2,43,44]. In general, $\Delta_{TSS}$ of a surface state is smaller than the SC gap of bulk bands ($\Delta_b$), due to the proximitized pairing amplitude decays from bulk to surface. Interestingly, a special candidate of connate TSC, $\beta$-Bi$_2$Pd film, may break this rule [34–36]. A previous scanning tunneling microscopy/spectroscopy (STM/S) experiment found two SC gaps ($\Delta_1$ ~1.0 meV and $\Delta_2$ ~3.3 meV) in the $\beta$-Bi$_2$Pd film grown by molecular beam epitaxy [35], while only the smaller one ($\Delta_1$) compares to the SC gap of $\beta$-Bi$_2$Pd bulk single crystal ($\Delta_b$ ~ 0.8 meV, $T_c$ = 5.4 K) [36,57,58]. Large zero-bias conductance peaks (ZBCPs) were observed in the line-cut measurement across its SC vortices. The ZBCPs do not split within a certain length from the vortex center, which indicates certain mixtures of MZMs inside the intensity of ZBCPs [35]. Consequently, the anomalous large gap ($\Delta_2$) was attributed to the enhanced $\Delta_{TSS}$ of the topological surface state, but the direct momentum-resolving evidence is still absent [35,59]. In this work, we performed angle-resolved photoemission spectroscopy (ARPES) measurements on as-grown $\beta$-Bi$_2$Pd thin film to directly resolve the origin of the large gap $\Delta_2$ [35]. We found the experimental evidences of an anomalously large SC gap at the Fermi-momentum ($k_F$) of topological surface state which likely corresponds to $\Delta_2$. On the contrary, no such large SC gap can be found in neither the trivial surface state nor the bulk bands. By comparing with bulk single crystal, we showed that the chemical potential is shifted upward in the film, which might be the cause for the deviation between different types of samples.

The (001)-oriented 20-UC $\beta$-Bi$_2$Pd thin films were epitaxially grown on Nb-doped (0.7 wt%) SrTiO$_3$(001) substrates at ~ 320 °C. High-purity Bi (99.9999%) and Pd (99.99%) sources were co-evaporated from Knudsen cells with a flux ratio of 5.3, which were measured by a quartz crystal monitor. Films were studied *in-situ* using home-made room-temperature STM and low-temperature ARPES with ultrahigh vacuum better than $3 \times 10^{-11}$ torr. The ARPES system is equipped with a Scienta R4000 analyzer and a helium discharge lamp with He I$\alpha$ photons (21.218 eV). The energy resolution was set ~ 3 meV for gap measurements and ~ 7 meV for band structure measurements. The angular resolution was set to ~0.2°. ARPES measurements on $\beta$-Bi$_2$Pd bulk single crystals with 20 eV photons were performed at the "Dreamline" beamline of the Shanghai Synchrotron Radiation Facility (SSRF) with a Scienta DA30 analyzer.

The 20-UC $\beta$-Bi$_2$Pd thin films measured in this work have tetragonal structure (space group *I4/mmm*) [Fig. 1(a)]. The lattice mismatch between substrate and $\beta$-Bi$_2$Pd is released as growing layer-by-layer. Lattice constants of thin film are $a = b = 3.41$ Å, $c = 12.97$ Å, obtained by *in-situ* reflection high-energy electron diffraction measurement [bottom panel of Fig. 1(d)] and X-ray diffraction (XRD) [Fig. 1(e)], which are in good agreement with the bulk single crystal. A STM image of the film [Fig. 1(c)] shows patch-like growing nature of $\beta$-Bi$_2$Pd, which is consistent with previous STM study that observed two SC gaps [35]. Our XRD measurements show (001)-oriented single crystallization of thin films. We performed ARPES measurements on these films with He-I$\alpha$ photons. Similar as the results of the bulk single crystal [34], four-fold symmetric Fermi surfaces with four Fermi pockets are resolved [Fig. 1(f)]. The band dispersion along $\bar{\Gamma}$-$\bar{X}$ [Fig. 1(b)] is plotted in Fig. 1(g), with two hole-like bands ($\alpha$, $\beta$) and one electron-like band ($\gamma$) crossing the Fermi level ($E_F$).

It has been resolved in $\beta$-Bi$_2$Pd bulk single crystal that a surface Dirac cone appears beneath the $\alpha$ band [34], with the binding energy of the Dirac point around -2.4 eV. It is known that in a thin film of TI within only a few layers, the topological surface states on the two sides may strongly hybridize with each other, leading to gap opening at the Dirac point [60,61]. An ideal TI preserving topological protection should be free of such a hybridization gap. We checked the spectra of high binding energy between -1.4 eV and -3 eV in our measurements. A clear Dirac dispersion [Fig. 2(a)] with isotropic constant-energy contours [Fig. 2(b)] can be observed, suggesting that our 20-UC thin films keep the topological surface states intact and is similar to the bulk material [34].

Next, we turn to the surface states near $E_F$. We display ARPES dispersion near $E_F$ along $\bar{\Gamma}$ - $\bar{X}$ [Fig. 2(c)]. Besides the bulk bands mentioned before [Fig. 1(g)], there are two distinguishable surface states observed in our measurements. With assistance from our first principle calculation (Fig. 4b), we clearly identify those bands as a trivial surface state ($\beta'$) deriving from the $\beta$ band and a topological surface state ($\gamma'$) connecting the $\beta$ and $\gamma$ bands. Remarkably, there is an obvious dip between $\gamma'$ and $\gamma$ bands in the momentum distribution curve (MDC) extracted at $E_F$ [the red curve appended in Fig. 2(c)], which is more distinct in the film as comparing to the previous study on the bulk single crystal [34]. The separation between the topological surface states and the bulk bands is clearly demonstrated in a curvature intensity plot around the $\gamma'$ and $\gamma$ bands [Fig. 2(d)]. It leads us to conjecture that more surface state components are presented in the film materials, which preserves the topological properties from overlapping with other bulk signals. We display five cuts along $k_y$ [Fig. 2(e)], with its $k_x$ positions indicated in Fig. 1(f). The Fermi-level MDCs show $\gamma'$ gradually merge into $\gamma$, when moving from the Brillouin zone center (cut#1) to the edge (cut#5) [Fig. 2(f)]. It implies that the surface components reach the maximum at $\bar{\Gamma}$-$\bar{X}$, which is the best place to study the intrinsic superconductivity of topological surface state ($\gamma'$) in the films.

Next, we focus on the momentum-resolved superconductivity of $\beta$-Bi$_2$Pd film. We performed high-resolution ARPES measurements along $\bar{\Gamma}$-$\bar{X}$ under different temperatures, *i. e.*, below $T_c$ (2.7 K) [Fig. 3(a)] and above $T_c$ (20 K) [Fig. 3(b)]. We notice that the topological surface state ($\gamma'$) bends toward higher binding energy when $k$ approaches $k_F$ at 2.7 K [Fig. 3(a)], while the band straightly crosses $E_F$ at 20 K [Fig. 3(b)]. This behavior implies the formation of a SC gap. The bending back feature is characteristics of Bogoliubov dispersion of SC state. The Bogoliubov quasiparticles produce a sharp coherent peak and its position can be defined as the size of SC gap [62–68]. We extracted energy distribution curves (EDCs) at three representative momenta, namely $k_1$, $k_2$ and $k_3$ [as marked in Fig. 2(c)], which correspond to the $k_F$ values of topological surface state ($\gamma'$), trivial surface state ($\beta'$) and bulk state ($\alpha$), respectively. Surprisingly, at $k_1$, the EDC measured at 2.7 K shows a sharp peak at -3.8 meV [Fig. 3(d)], which was in contrast with the featureless EDC measured at 20 K [Fig. 3(d)]. We attributed this sharp peak as the enhanced SC gap as measured in the previous STM/S study ($\Delta_2 \sim 3.3$ meV) [35]. These observations were reproduced several times in different samples [69], which strengthens us confidence of the existence of SC topological surface states in $\beta$-Bi$_2$Pd films with an anomalously large SC gap. We notice that the EDCs measured on the $k_F$ of trivial surface state [Fig. 3(e)] and bulk band [Fig. 3(f)] are featureless near $E_F$, even at 2.7 K. This observation is reasonable because the SC gap values of those bands ($\Delta_1 \sim 1$ meV of films [35] and $\Delta_b \sim 0.8$ meV of bulk single crystal [36,57]) [Fig. 3(c)] are much smaller than the experimental energy resolution of our ARPES system (~ 3 meV). We summarize the gap sizes measured by different techniques in Fig. 3(c), the SC gap measured in this work are comparable to previous STM/S observations [35]. The appearance of two classes of SC gaps indicates the paring potential is indeed enhanced on the topological surface states.

In order to resolve the puzzle of the anomalous SC gap enhancement on the topological surface state in this film material, we conducted comparison studies between 20-UC films and bulk single crystals of $\beta$-Bi$_2$Pd. We observed that the chemical potential shifts upward about 37meV in the thin film [Fig. 4(a)]. We repeated this measurement for several times on different samples and obtained confirming results [69]. Theoretically, the odd and even components of the SC order parameter can mix with each other on the sample surface due to inversion symmetry broken [70]. A similar phenomenon of enhanced $\Delta_{TSS}$ was proposed in Cu$_x$Bi$_2$Se$_3$ previously [71], that the orbital polarization of topological surface states leads to constructive parity mixing of SC order parameters. However, the trivial surface states cannot support such constructive mixing, although odd and even components of the order parameter do coexist on the surface [71,72]. It was suggested that a larger Fermi momentum separation between topological surface states and adjacent bulk band ($\delta k$), equivalently a larger surface weight of topological surface states, can lead to a stronger enhancement of $\Delta_{TSS}$ [71]. However, the $\delta k$ difference between thin film and bulk single crystal is not quite clear in our experiment. So that we performed a slab calculation to simulate the surface weight of topological surface states at different chemical potentials [69]. The calculated band structure is consistent with our experimental results and previous studies [34,73]. The color scale in Fig. 4(b) indicates the surface weight. Apparently, the surface weight becomes larger when the chemical potential is increased [inset of Fig. 4(b)]. Although our calculation qualitatively supports the mechanism of Dirac-Fermion-mediated parity mixing [71] in explaining $\Delta_{TSS}$ enhancement, the surface weight difference between bulk single crystal and thin film is only ~ 6%. Thus we caution on how such a small change could lead to this large enhancement of SC gap. A realistic model or a different mechanism may be needed in order to resolve this puzzle.

In conclusion, we performed *in-situ* ARPES measurements on $\beta$-Bi$_2$Pd films and bulk single crystals. We observed a direct momentum-resolved evidence of an anomalously large SC gap on its topological surface state. A possible enhancing mechanism, which is the Dirac-Fermion-mediated parity mixing [71], was discussed based on our observation of chemical potential shift and the concomitant increasing of surface weight revealed in our first principle calculations.

**Acknowledgements:**

We thank C. Fang for useful discussions. This work at IOP is supported by the grants from the Ministry of Science and Technology of China (2016YFA0401000, 2016YFA0300600, 2015CB921000), the Natural Science Foundation of China (11888101, 11574371, 11622435, 11474340, 11774399), the Chinese Academy of Sciences (XDB28000000, XDB07000000, QYZDB-SSW-SLH043), and the Beijing Municipal Science and Technology Commission (Z171100002017018, Z181100004218005, Z171100002017018). Beijing Natural Science Foundation (Z180008), the National Key Research and Development Program of China (2017YFA0302901), Y.-B. H. acknowledges supports by the Ministry of Science and Technology of China (2016YFA0401002) and the CAS Pioneer "Hundred Talents Program" (type C). H.D. and Y.-J. S designed the experiments and supervised the project. J.-Y. G. and Y.-J. S. grew the thin films, performed STM, RHEED and XRD measurements. L.-Y. K. and J.-Y. G. performed ARPES measurements with the assistance of H. Li., Y.-G. Z., H.-J. L., C.-Y. T., F.-Z. Y., Y.-B. H. and T. Q.. D.-Y. Y. and Y.-G. S. provided high quality bulk single crystals. L.-Q. Z. and H.-M. W performed first principle calculation. J.-Y. G. and L.-Y. K. analysis the ARPES data. J.-Y. G. plotted the figures. L.-Y. K., J.-Y. G., Y.-J. S. and H. D. wrote the manuscripts with inputs from all the authors.

[*] These authors contributed equally to this work.

[†] Corresponding authors.
  dingh@iphy.ac.cn
[‡] Corresponding authors.
  yjsun@iphy.ac.cn

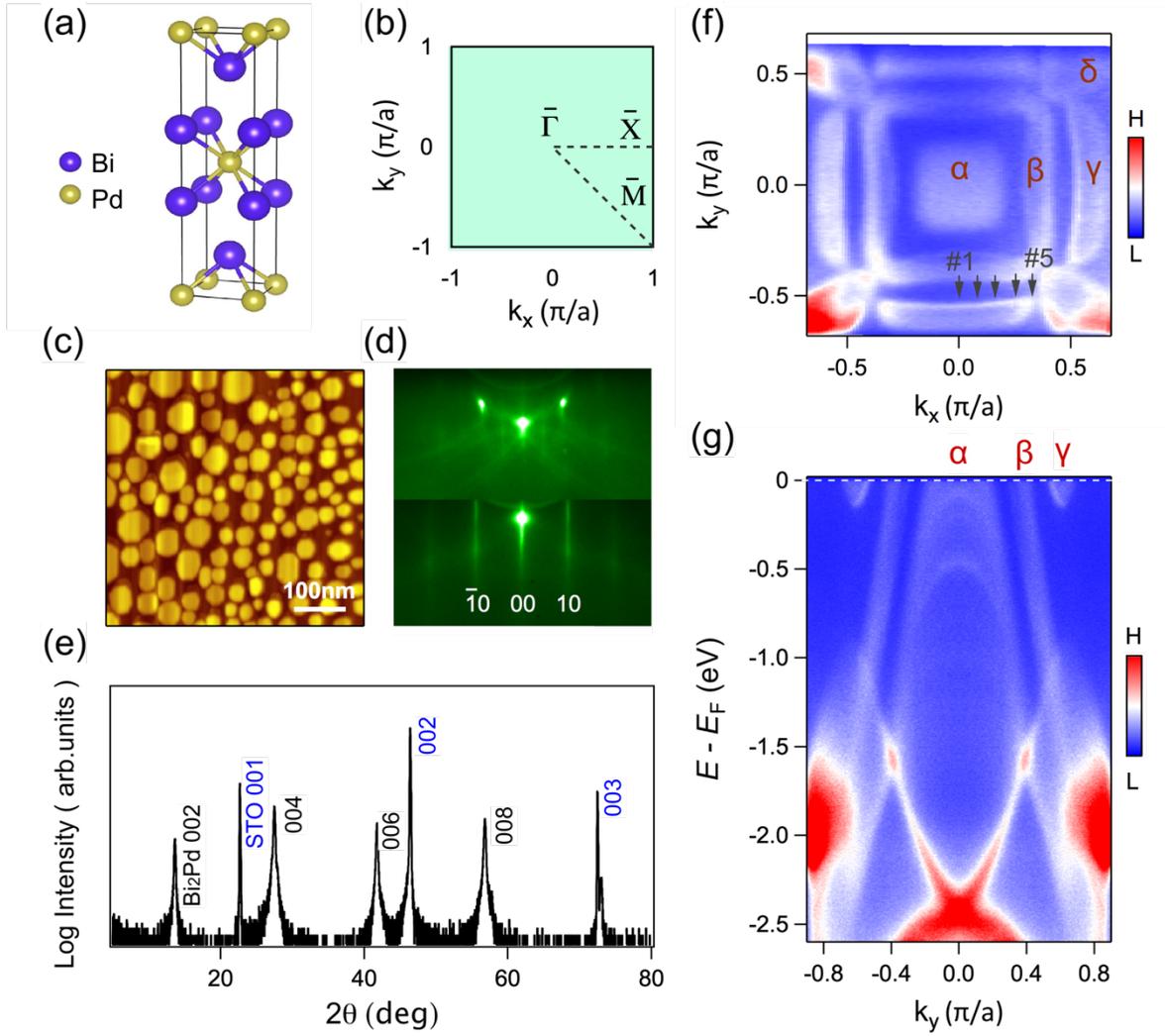

FIG. 1. (a) Crystal structure of tetragonal $\beta$-Bi$_2$Pd film. Layers are stacked by van der Waals interaction. Each unit cell (UC) is made up of two Bi-Pd-Bi triple layers. (b) Projected surface Brillouin zone with high symmetry points ($\bar{\Gamma}$, $\bar{M}$ and $\bar{X}$). (c) Constant-current STM topographic image of as-grown 20 UC $\beta$-Bi$_2$Pd film (setpoint voltage: $V_s$ = 2.13 V, tunneling current: $I_t$ = 270 pA, 500 nm × 500 nm). (d) Reflection high-energy electron diffraction pattern taken from the (001) surface on an annealed SrTiO$_3$ substrate (top panel) and that of $\beta$-Bi$_2$Pd film with Kikuchi lines formed by inelastically scattered electrons (bottom panel), indicating high crystalline coherence. (e) X-ray diffraction spectrum taken from the same film illustrates lattice constant $c$ = 12.97 Å (X-rays with wavelength 1.54 Å). (f) Four-fold symmetrized Fermi surface obtained by ARPES at 20 K shows spectral weight within $E_F$ ±10 meV. The Fermi surface is composed of two hole bands ($\alpha$, $\beta$) and two electron bands ($\gamma$, $\delta$). Black arrows with numbers #1 to #5 mark the positions of the cuts shown in Fig. 2(e). (g) Large-range ARPES spectrum observed along $\bar{\Gamma}$ - $\bar{X}$.

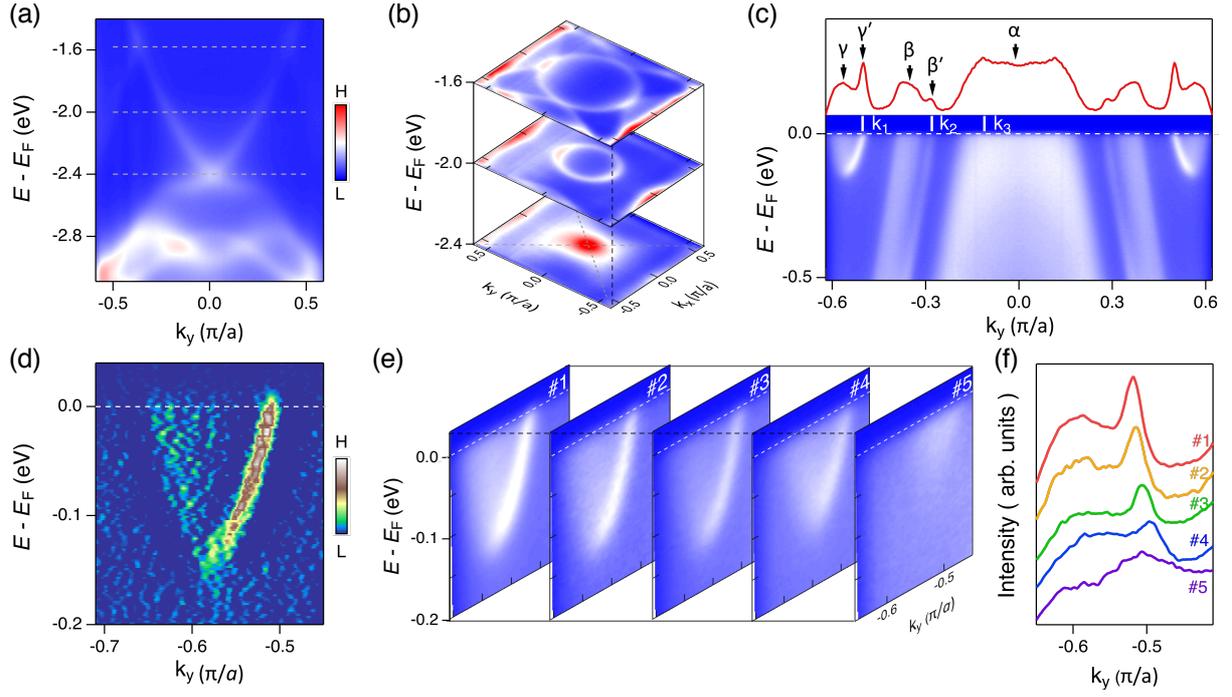

FIG. 2. (a) High binding energy band dispersion of 20-UC $\beta$-Bi$_2$Pd film. An intact surface Dirac cone disperses along $\bar{\Gamma}$ - $\bar{X}$. (b) Constant energy contours at binding energy $E_D$, $E_D$+0.4 eV and $E_D$+0.8 eV, where $E_D$ (-2.4 eV) is the energy of the Dirac point. (c) Close-up of ARPES spectrum (along $\bar{\Gamma}$ - $\bar{X}$) near $E_F$ measured at 20 K. The topological surface state $\gamma'$ connects the $\gamma$ and $\beta$ bulk states, and the trivial surface state $\beta'$ derived from the bulk state $\beta$. The red line represents the extracted momentum distribution curve (MDC) at $E_F$. Three representative momenta, namely $k_1$, $k_2$ and $k_3$, correspond to the Fermi momentum of $\gamma'$, $\beta'$ and $\alpha$ bands, respectively. (d) Curvature intensity plot of the $\gamma$ and $\gamma'$ bands. (e) Momentum dependence of the $\gamma$ and $\gamma'$ dispersions, $k_x$ positions of cuts #1 to #5 are indicated in Fig. 1(f). (f) MDCs extracted at $E_F$ for the five cuts, offset for clarity.

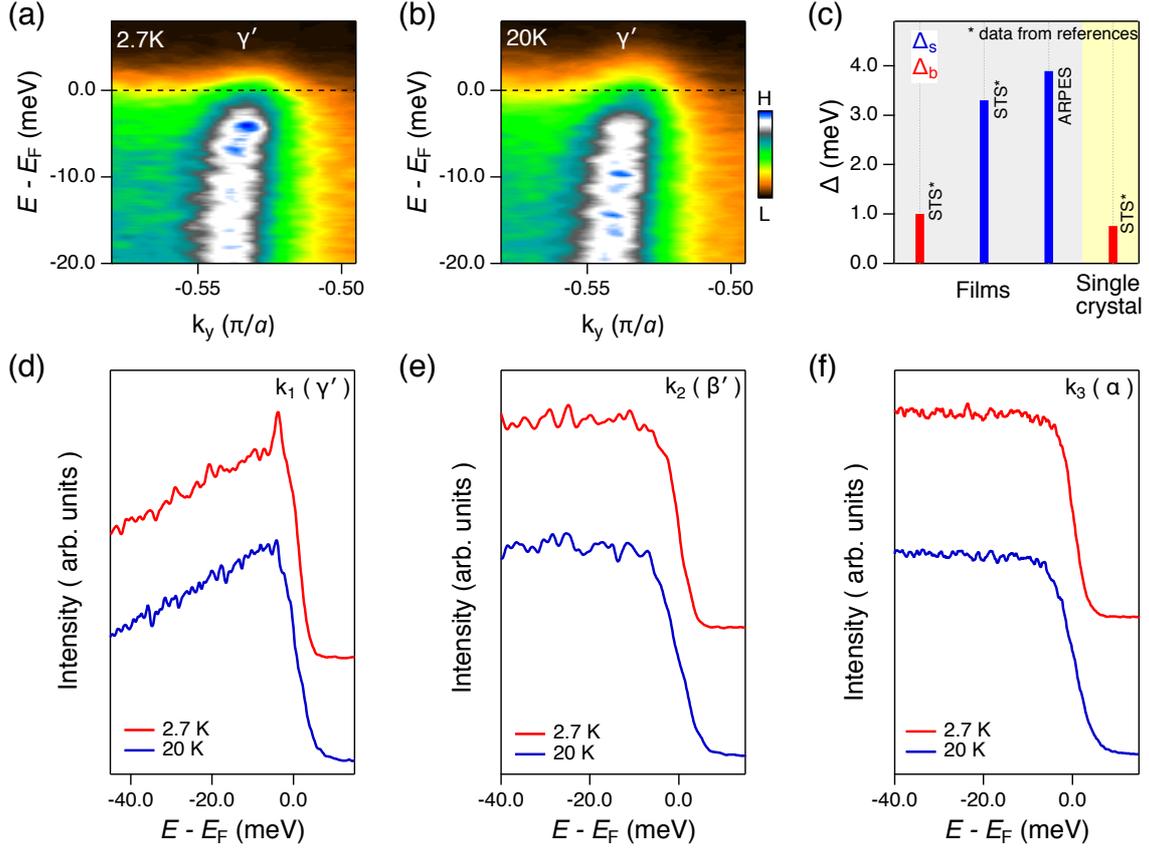

FIG. 3. Close-up of ARPES spectra near $E_F$ along $\bar{\Gamma}$ - $\bar{X}$ measured on the topological surface state ($\gamma'$) at 2.7 K (a) and 20 K (b), respectively. (c) A collection of SC gaps measured on $\beta$-Bi$_2$Pd samples. Red (blue) color represents the size of SC gap on the topological surface state $\Delta_s$ (bulk band $\Delta_b$). The SC gap measured on bulk single crystal is about 0.8 meV measured by scanning tunneling spectroscopy (STS) [36,57]. The SC gap values measured on the film in a previous STM/experiments [35] are 1 meV ($\Delta_1$) and 3.3 meV ($\Delta_2$), which are assigned as SC gap on bulk bands and topological surface states, respectively. The SC gap measured on the topological surface state in this work is 3.8 meV. (d) The energy distribution curves (EDCs) are extracted on the momentum $k_1$ (topological surface state) at 2.7 K (red curve) and 20 K (blue curve). (e) and (f) same as (d) but measured on momenta $k_2$ (trivial surface state) and $k_3$ (bulk state), respectively.

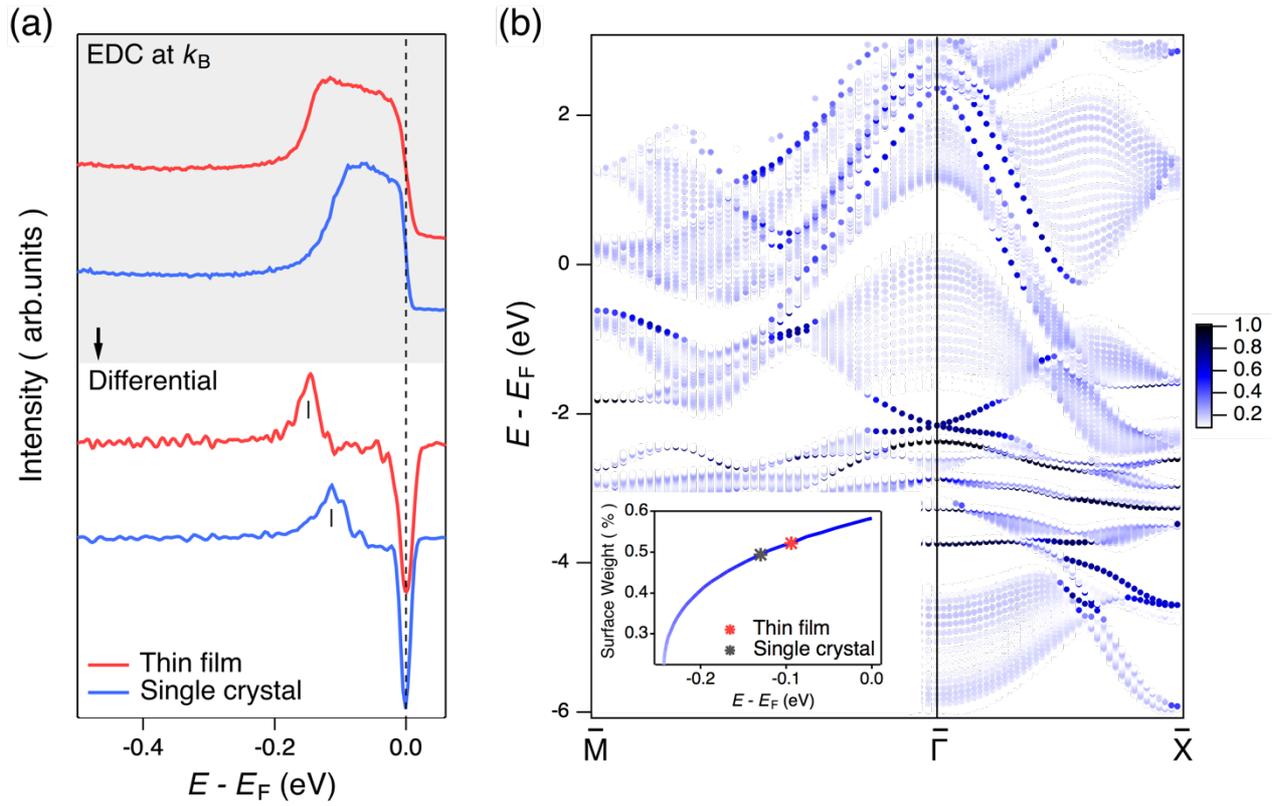

FIG. 4. (a) EDCs at $k_B$ taken from a bulk single crystal (blue curve) and a film (red curve) where $k_B$ represents the momentum of the $\gamma$ band bottom (top panel). We define the binding energy of the band bottom by the peaks of first derivative of the EDCs (bottom panel). The chemical potential of the thin film shifts upward ~37 meV as comparing with the bulk single crystal. (b) The projection of (001) surface bands obtained by slab calculation of 11 $Bi_2Pd$ layers. The color scale indicates the weight of surface component. The inset shows the surface weight of $\gamma'$ (blue curve). The position of the chemical potential for the thin film (the bulk single crystal) as marked by red (grey) dot.

# Supplemental Material for
# Experimental evidence of anomalously large superconducting gap on topological surface state of *β*-Bi₂Pd film

J.-Y. Guan[1,2,*], L.-Y. Kong[1,2,*], L.-Q. Zhou[1,2], Y.-G. Zhong[1,2], H. Li[1,2], H.-J. Liu[1,2], C.-Y. Tang[1,2], D.-Y. Yan[1,2], F.-Z. Yang[1,2], Y.-B. Huang[3], Y.-G. Shi[1,5], T. Qian[1,4,5], H.-M. Weng[1,5], Y.-J. Sun[1,4,5,‡] and H. Ding[1,4,5,†]

[1] *Beijing National Laboratory for Condensed Matter Physics and Institute of Physics, Chinese Academy of Sciences, Beijing 100190, China*
[2] *School of Physics, University of Chinese Academy of Sciences, Beijing 100190, China*
[3] *Shanghai Synchrotron Radiation Facility, Shanghai Institute of Applied Physics, Chinese Academy of Sciences, Shanghai 201204, China*
[4] *CAS Center for Excellence in Topological Quantum Computation, University of Chinese Academy of Sciences, Beijing 100190, China*
[5] *Songshan Lake Materials Laboratory, Dongguan, Guangdong 523808, China*

## I. Film flatness characterization and single crystal growth

Scanning tunneling microscopy (STM) is an ideal technique for studying sample surface. Here we show a topographic image of *β*-Bi₂Pd film measured by STM, then deduce the height profile from the green line marked in Fig. S1(a). The patches height on outmost layer is ~ 3.8 nm [Fig. S1(b)], comparable with the height of 3-UC. The film surface is flat, and the medium thickness of *β*-Bi₂Pd film is 3 unit-cell in Fig. S1(a). The outmost 3-UC patches grew on intact inner layers of *β*-Bi₂Pd film.

Bulk single crystals of *β*-Bi₂Pd were grown using a self-flux method. The starting materials were mixed in a molar ratio of 1:2 in a glovebox filled by argon. The mixture was placed in an alumina crucible sealed in a fully evacuated quartz tube. The crucible was heated to 900 ℃ in 10 h. After dwelled for 24 h, it was cooled slowly to 430 ℃ at 2 K/h. Single crystals were gained after quenched in the cold water. One of single crystal samples is shown in Fig. S1(c).

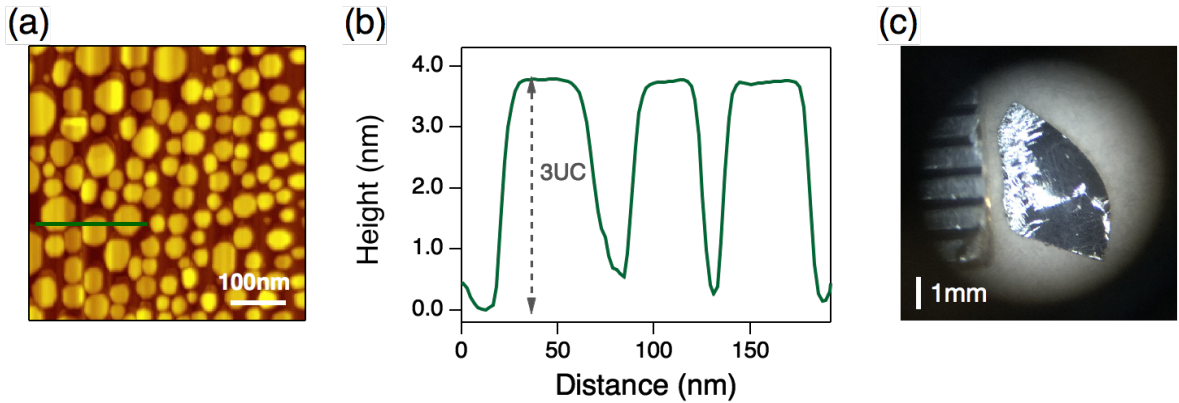

FIG.S1. Film flatness characterization and single crystal growth. (a) Constant-current STM topographic image of as-grown ~20UC thick *β*-Bi₂Pd film ($V_s$ = 2.13 V, $I_t$ = 270 pA, 500 nm × 500 nm). (b) The height profile deduced from the green line marked in (a). (c) The picture taken by microscopy on *β*-Bi₂Pd single crystal used in our ARPES measurements.

## II. First principle calculation

We constructed 11 layers of supercell to realize the projection of surface bands based on the density functional theory(DFT), and added 12Å thick vacuum layer along the (001) direction to avoid the interaction between slabs. SOC was taken into account. The calculation was implemented in the Vienna Ab-initio Simulation Package(VASP) [S1] and the generalized gradient approximation (GGA) with the Perdew-Burke-Ernzerhof (PBE) [S2] functional were selected to describe the exchange-correlation energy. The energy cutoff value of 500 eV for planewave basis was adopted. The reciprocal space was sampled by 10×10×1 k-point grids in the Brillouin zone (BZ) for the slab and the optimal force convergence threshold was set as 0.001eV/Å on each atom. The result is shown at Fig. 4(b). We select all the atoms of the top and bottom unit cell layer of the 11-layer supercell as the surface atoms. The orbital projections of these 12 surface atoms are called the surface weight.

To further compare the (001) surface state of $Bi_2Pd$, we constructed the maximally-localized Wannier function (MLWF) for Bi 6p orbital and Pd 4d orbital using the method developed by Marzari and co-workers [S3]. Then the semi-infinite Green function method was used to calculate the surface state [S4]. The result is shown at Fig. S2.

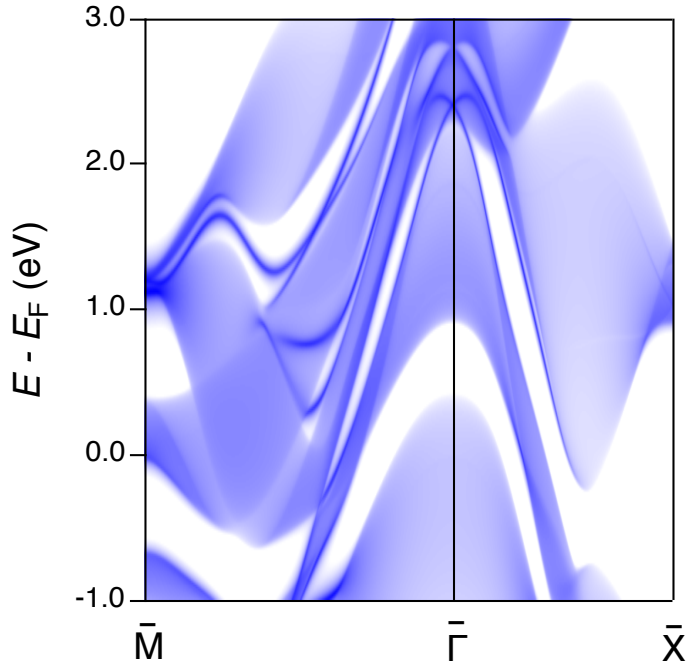

FIG. S2. Surface and bulk band dispersions obtained by tight-binding approximation.

## III. The reproducibility of superconductivity on Dirac surface state

Fig. S3(a, b) show the wide range dispersions of nontrivial surface state on $\beta$-Bi$_2$Pd film sample #1 which is described in main body. Besides, we measured the superconductivity of topological surface state on different as-grown films as well. The data is reproducible. Here we display the feature of bending back band on thin film sample #2 at low temperature [Fig. S3(c)]. And the nontrivial surface state band straightly crosses the $E_F$ at high temperature [Fig. S3(d)].

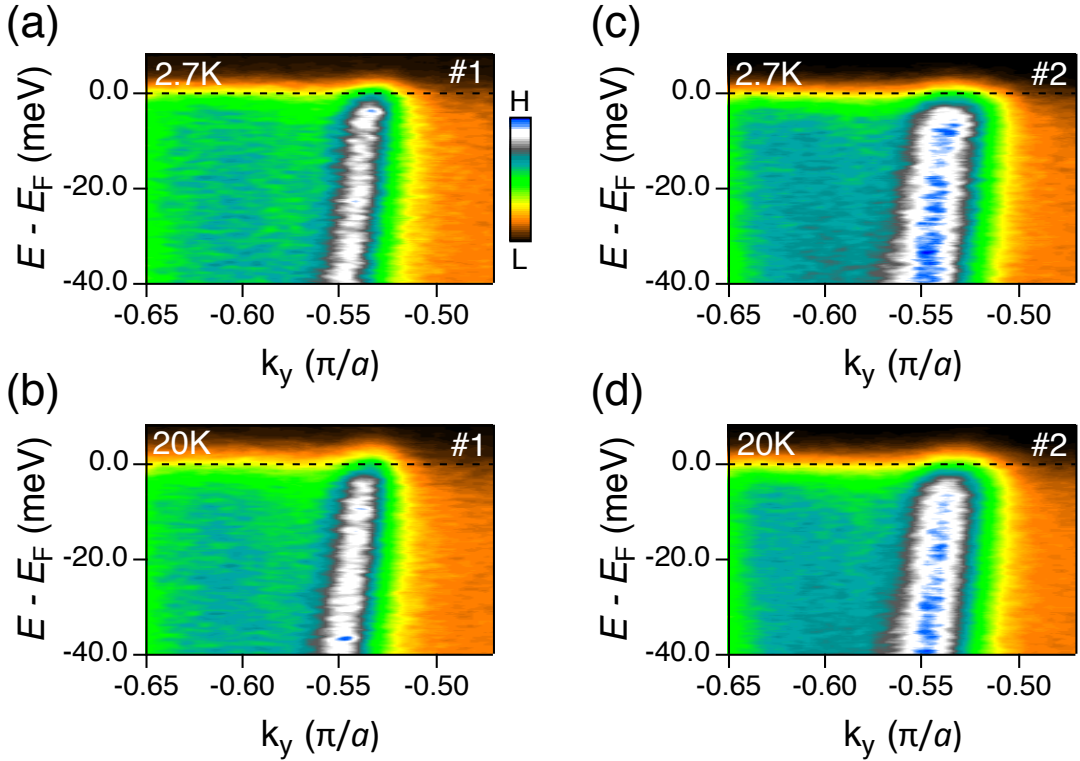

FIG.S3. The reproducible superconductivity of topological surface state. (a) Close-up of ARPES spectra near $E_F$ along $\bar{\Gamma}$ - $\bar{X}$ cut measured on nontrivial surface state ( $\gamma'$ ) of thin film sample #1 at temperature 2.7 K and 20 K (b). Sample #1 is described in main body. (c) and (d) same as (a) and (b) but measured on thin film sample #2, respectively.

## IV. The reproducibility of upward-shift of chemical potential in $\beta$-Bi$_2$Pd film

In order to visualize the shift of chemical potential, we display ARPES spectra of both film and bulk single crystal [Fig. S3(a, b), respectively]. The shift of chemical potential is very obvious in zoom-in plots of $\gamma$ and $\gamma'$ bands [Fig. S3(c)]. Then, we listed the enhanced chemical potential of different films [Fig. S3(d)]. The upward-shift of chemical potential in films is well reproducible.

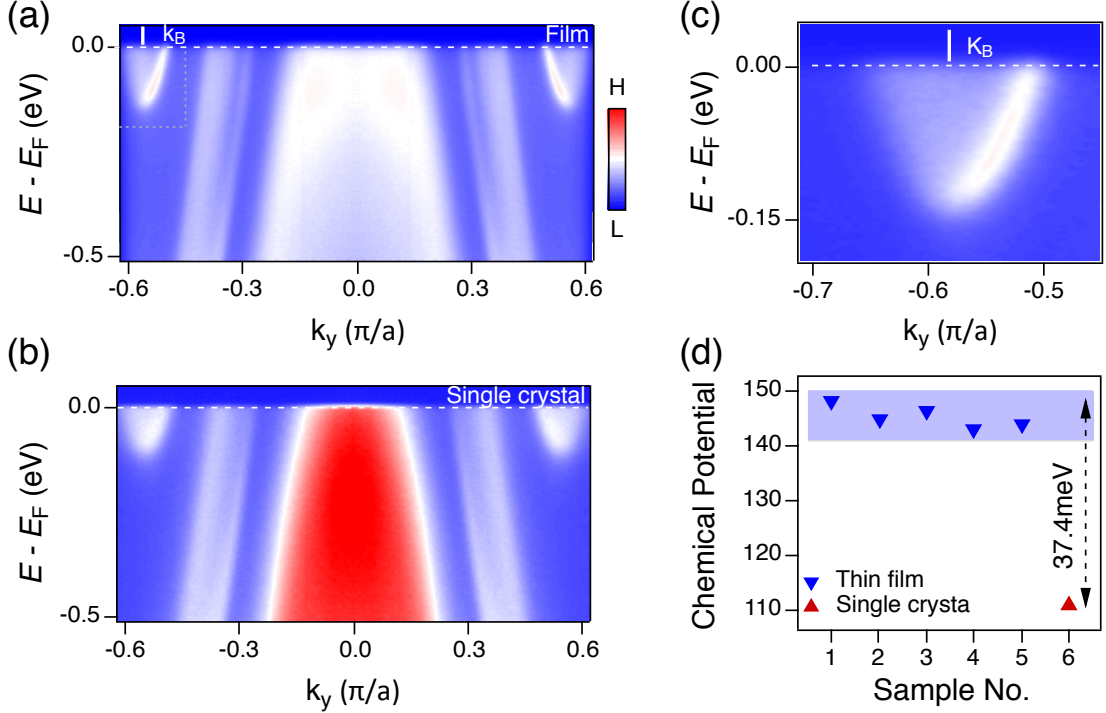

FIG.S4. The reproducibility of upward-shift of chemical potential in thin film. (a) Close-up of ARPES spectrum along $\bar{\Gamma}$ - $\bar{X}$ near $E_F$ measured at 20 K in thin film. The momentum $k_B$ corresponds to the bottom of $\gamma$. (b) same as (a) but measured on single crystal at 20 eV. (c) Amplified dispersions at rectangular region as shown in (a). (d) The chemical potential of different thin films and single crystal $\beta$-Bi$_2$Pd samples.